\documentstyle[twocolumn,twoside,prl,aps,epsfig]{revtex}

\begin{document}
\title{Interface resistance of disordered magnetic multilayers }
\author{K. Xia$^{1}$, P. J. Kelly$^{1}$, G. E. W. Bauer$^{2}$,
I. Turek$^{3}$, J. Kudrnovsk\'y$^{4}$, and V. Drchal$^{4}$}
\address{$^{1}$Faculty of Applied Physics and MESA$^{+}$ Research
Institute,
University of
Twente,\\ P.O. Box 217, 7500 AE Enschede, The Netherlands\\
$^{2}$Department of Applied Physics and DIMES, Delft University of
Technology,\\
Lorentzweg 1, 2628 CJ Delft, The Netherlands\\
$^{3}$ Institute of Physics of Materials, Academy of Sciences of the
Czech Republic, CZ-616 62 Brno, Czech Republic \\
$^{4}$ Institute of Physics, Academy of Sciences of the Czech
Republic, CZ-182 21 Prague, Czech Republic}
\date{\today}

\maketitle

\begin{abstract}
We study the effect of interface disorder on the spin-dependent interface
resistances of Co/Cu, Fe/Cr and Au/Ag multilayers using a newly
developed method for calculating transmission matrices from
first-principles.
The efficient implementation using tight-binding linear-muffin-tin
orbitals  allows us to model interface disorder using large lateral
supercells
whereby specular and diffuse scattering are treated on an equal
footing.  Without introducing any free parameters, quantitative agreement
with 
experiment is obtained. We predict that disorder {\it reduces} the
majority-spin interface resistance of Fe/Cr(100) multilayers by a factor 3.
\end{abstract}



When two layers of magnetic material are separated by a non-magnetic
spacer layer, the electrical resistance of the system depends strongly
on whether the magnetization directions are aligned parallel or
anti-parallel. This effect is known as giant magnetoresistance
(GMR)\cite{Baibich88}.
The huge interest\cite{Levy,GijsBauer,Barthelemy99} in the physics of
GMR is largely driven by the wide application potential of the effect,
which has already been realized in magnetic recording heads.

GMR can be observed in a number of different measuring
configurations.  The current-in-plane (CIP) configuration is experimentally
the 
simplest and is what is used at present in applications. However, for
gaining a better understanding of the underlying physics, the
current-perpendicular-to-the-plane (CPP) configuration
\cite{GijsBauer,Pratt91,Gijs93,Valet,Bass99} is preferred because of its
higher symmetry which should make it easier to understand and because of
higher MR ratios. 

The factors usually considered in theoretical treatments of GMR are
the
potential steps encountered by electrons passing from one material to
another, impurity scattering in the bulk of the layers, and defect
scattering at the interfaces\cite{Levy,GijsBauer,Barthelemy99}. There
has
been a great deal of discussion about the relative importance of these
ingredients and their spin dependence, which cannot be resolved solely
on the basis of model calculations which include these effects in
parameterized form. Once the question has been suitably posed,
however, 
detailed electronic structure calculations can be used to resolve the
issue quantitatively. For example, the effect of potential steps
and their microscopic origin could be established in this
way\cite{Schep95,Zahn95}.

In this Letter we wish to address the relative role
of specular and diffuse interface scattering. This has been studied by
a large number of authors but so far only using simple models which do
not allow for detailed quantitative analysis of specific
materials\cite{Brataas94,Zhang98}. We focus
on the interface resistance of the resistor model which describes the
observed thickness and layer dependence of CPP-GMR remarkably
well\cite{GijsBauer}. Because it turns out to be strongly spin-dependent and
dominates the magnetoresistance for layer thicknesses which are not
too large, the key to understanding CPP
magnetoresistance lies in understanding the origin of the interface
resistance.  The methodology which we have developed allows us to
include specular and diffuse scattering on an equal footing without
introducing any arbitrary fitting parameters.


Explicit expressions for the interface resistance were derived by
Schep {\em et al.}~\cite{Schep97}
in terms of the transmission matrix $T$ which describes how the electronic
structure mismatch at an $A/B$ interface affects electron transport. In the
limit in
which there is no coherent scattering between adjacent interfaces,
presumably 
due to sufficiently strong bulk scattering, the
interface resistance is given by
\begin{equation}
R_{A/B}=\frac{h}{e^{2}}\left[ \frac{1}{\sum T_{\mu \nu }}-
\frac{1}{2}\left( 
\frac{1}{N_{A}}+\frac{1}{N_{B}}\right) \right].
\label{Rint}
\end{equation}
where $T_{\mu \nu}$ are the probabilities for
eigenstate $\mu$ in material $A$ to be transmitted through the
interface into 
the eigenstate $\nu$ in material $B$, the sum is over the Fermi
surface,  and ${e^2 / h}\; N_{A(B)}$ is the Sharvin conductance of material
A(B). 
In Ref.~\cite{Schep97} transmission matrices and interface resistances were
obtained for ideal Co/Cu interfaces
using a first-principles FLAPW based embedding technique. These, and
similar results obtained by Stiles and Penn\cite{Stiles00},
demonstrated that a combination of spin-independent bulk scattering
and strongly spin-dependent specular interface scattering arising from the
spin-dependence of the band mismatch can account for the observed
spin-dependence of the interface resistances. These results are at odds with
the common wisdom that metallic heterointerfaces cannot be perfect
due to unavoidable roughness and/or interface alloying. Indeed, for the one
case
of Co/Cu(111) interfaces for which direct comparison could be made
with experiment, the agreement though reasonable, was not perfect. We
therefore address the following questions: Why does a calculation for a
perfect interface
agrees so well as it does for a sample produced by sputtering? Can theory
and experiment be
brought to even better agreement by taking into account disorder? Is the
finding that
specular interfaces are a reasonable first order approximation generic or a
coincidence
found only for the Co/Cu system?

The FLAPW-based method used in Ref.~\cite{Schep97} was computationally
too demanding to allow interface disorder to be treated. Starting
instead with the more efficient surface Green's function
method\cite{Turek} implemented with a tight-binding linear muffin tin
orbital basis\cite{Andersen85}, we can now calculate the transmission
and reflection matrices needed in the Landauer-B\"uttiker formulation
of transport theory\cite{Datta95}, but now for much larger systems.
In this Letter, we present the results of calculations for Co/Cu,
Fe/Cr and Ag/Au layered systems in which we model interface disorder by
means of large lateral supercells.
The electronic structure is determined self-consistently within the
local spin density approximation. To model the interface, we randomly
distribute the appropriate concentration of different atoms within
lateral supercells\cite{Bruno} containing as many as $10 \times 10$ atoms.
For the
disordered layers the potentials are determined self-consistently
using the layer CPA approximation\cite{Turek}.
The calculations are carried out with a ${\bf k_{\|}}$ mesh density
equivalent to 3600 ${\bf k_{\|}}$ mesh points in the two dimensional
Brillouin zone (BZ) of a $1 \times 1$ interface unit cell.
The numerical error bar resulting from this sampling is smaller
than 0.2\% of the conductance. The interface resistances calculated
for Co/Cu (100) and (111) in the clean limit using Eq.~(\ref{Rint})
agree with those obtained by Schep using an entirely different code
to within about ${\rm 0.1 ~f\Omega m^2 }$ or 5\%.

In the presence of defects, the conductance can be expressed as the
sum of a ballistic part and a diffuse part;
the transmission matrix elements between two Bloch states with the
same ${\bf k}_{\|}$ correspond to ballistic scattering, those between
two Bloch states with different ${\bf k}_{\|}$ to diffuse scattering.
%

\begin{table}[h]
\caption[Tab1]{Results of calculations}
\begin{tabular}{cccc}
system & roughness & ${\rm R_{maj}(f\Omega m^{2})}$ & ${\rm R_{\min }
(f\Omega m^{2})}$ \\
\hline
Au/Ag(111) & clean                & 0.094             & 0.094 \\
Au/Ag(111) & 2 layers 50-50 alloy & 0.118             & 0.118 \\
Au/Ag(111) & exp.                 & $0.100 \pm 0.008$ & $0.100 \pm
0.008$ \\ 
\hline
Co/Cu(100) & clean & 0.33 & 1.79 \\
Co$_{hcp}$/Cu(111) & clean & 0.60 & 2.24 \\
Co/Cu(111) & clean & 0.39 & 1.46 \\
Co/Cu(111) & 2 layers 50-50 alloy & 0.41 & 1.82$\pm 0.03$ \\
Co/Cu(111) & exp. & 0.26$\pm 0.06$ & 1.84$\pm 0.14$ \\
\hline
Fe/Cr(100) & clean       & 2.82 & 0.50 \\
Fe/Cr(100) & 2 layers 50-50 alloy & 0.99 & 0.50 \\
\end{tabular}
\label{tabone}
\end{table}

The calculated results are shown in Table~\ref{tabone}. The Au/Ag
interface has fcc(111) texture and the interface roughness is
estimated 
to be at least two layers thick in the MSU samples\cite{Henry96}.
This makes it very suitable for testing our method without the
complicating factor of spin dependence.
The interface resistance we find for a clean Au/Ag interface based on
Eq.~(\ref{Rint}) is
${\rm 0.047 ~f\Omega m^{2}}$ which is very close to the experimental
value, 
${\rm 0.050 \pm 0.004 ~f\Omega m^{2} }$.
The resistance of a Au/Ag interface becomes
${\rm 0.059 ~f\Omega m^{2} }$ when the interface contains two layers
of
Au$_{0.5}$Ag$_{0.5}$ alloy. The uncertainty arising from using
different 
alloy configurations within the $6 \times 6$ unit cell is less than
1\%.


We calculate the interface resistance of Co/Cu interfaces for both
(100) and (111) orientations. The lattice constant used for fcc Co/Cu
is 3.549~\AA . We focus mainly on the (111) orientation as this is
the structure which is predominantly seen in the experimental
samples. 
The interface alloy is again at least
two atomic layers thick\cite{Bass99}. We treat the interface disorder
as two layers of CoCu alloy modelled using an $8\times 8$ lateral supercell.
The largest uncertainty between different configurations of two layers
of 50-50 alloy is about $2.5\%,$ which is much smaller than the
experimental error bar.
For interface alloy compositions ranging from 50-50 to 44-56 the
interface resistance does not change within our numerical accuracy.
With two layers of interface alloy, the calculated transmission
probability for the minority spin electrons decreases by about 10\%
bringing the calculated interface resistance into near perfect
agreement with experiment.
We find that disorder gives rise to mainly forward scattering of the
electrons so that the decrease of the ballistic component is almost
cancelled by the increase of the diffusive part.
This is the reason why the calculations for the defect-free Co/Cu
interface \cite{Schep97,Stiles00} were in reasonable agreement with
experiment. The strong diffuse scattering also explains why the two-channel
resistor model
performs so well down to relatively thin layers in which bulk scattering
should not be
important.

The resistance of a Cu/Co interface calculated with two layers of
50\% interface alloy is
${\rm 0.41 ~f\Omega m^2 }$ for the majority spin and
${\rm 1.82 \pm 0.03 ~f\Omega m^{2} }$ for the minority spin.
The error bar for the minority-spin results from using a finite
lateral supercell for modelling disorder and configuration averaging.
The majority spin bands of Cu and Co, like the bands of Au and Ag,
are so well matched that interface disorder has very little effect
on the interface resistance.
We observe (Table I) that there is near perfect agreement for the
minority spin (certainly within the overall uncertainty of the
calculation) but that the calculated resistance for the majority spin
case which was already too large in the absence of disorder is even
slightly increased by disorder.

\begin{figure}[t]
\epsfxsize=6cm
\epsfbox{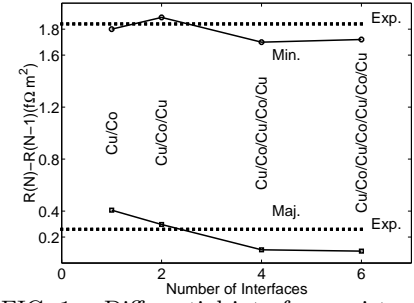}
\caption{
Differential interface resistance as the number of interfaces
increase for a disordered Cu/Co multilayer embedded between Cu
leads. }
\label{Fig1}
\end{figure}

With two layers of interface alloy, almost 80\% of the minority spin
current results from diffuse scattering.
We can see this in a different way by calculating the conductance for
Cu/Co/Cu, Cu/Co/Cu/Co/Cu, and Cu/Co/Cu/Co/Cu/Co/Cu in a manner quite
analogous to that described above for the single Cu/Co interface.
To be able to perform this large calculation we had to use a smaller
supercell ($6\times 6$) so that the error bar is larger than for the
$8 \times 8$ calculation.
In these calculations the boundary Cu layers are semiinfinite
``leads'' and the interface disorder is two layers of 50-50 alloy. Results are insensitive to the
individial layer thicknesses, chosen here to be 10 atomic layers. In Fig.~\ref{Fig1} the
differential resistance per interface is shown for all four systems.
For the minority spins, the deviation from the result obtained for
a single interface is quite small.
The strong diffuse interface scattering destroys the phase coherent
scattering between
subsequent  interfaces so that Ohm's law holds when the number of interfaces
is
increased. For the majority spin, however, the interface resistance does not
obey
Ohm's law but decreases as the number of interfaces increases.
It appears to saturate at a value of ${\rm 0.07 ~f\Omega m^2 }$ which
is only a third of the experimental value, but consistent with
Mathon's calculation for a multilayer with random layer
thicknesses\cite{Mathon}.
The majority spin potentials are so similar that the scattering from
a double alloy layer is insufficient to break the coherence which is
considerably longer range than for the strongly scattered minority
spins. 
We would have to assume that the majority spin electrons remain
coherent for transport through 4 interfaces in order to obtain a
value of the interface resistance close to the experimental value of
${\rm 0.26 ~f\Omega m^{2} }$.
Compared to real samples with bulk defects and lateral variations in
the layer thicknesses, it appears that we overestimate the coherence
length in the majority spin case\cite{Barthelemy99}.

For the (111) orientation we also considered an interface between
hcp Co and fcc Cu. 
For a clean 
${\rm Co^{(0001)}_{hcp}/ Cu^{(111)}_{fcc} }$ interface both majority
and minority spin resistances are substantially larger than for the
fcc case and even larger than the experimental values
(Table~\ref{tabone}).
In view of the substantial difference predicted between the two,
it should be interesting to try and measure it.

\begin{figure}[h]
\epsfxsize=9cm
\epsfbox{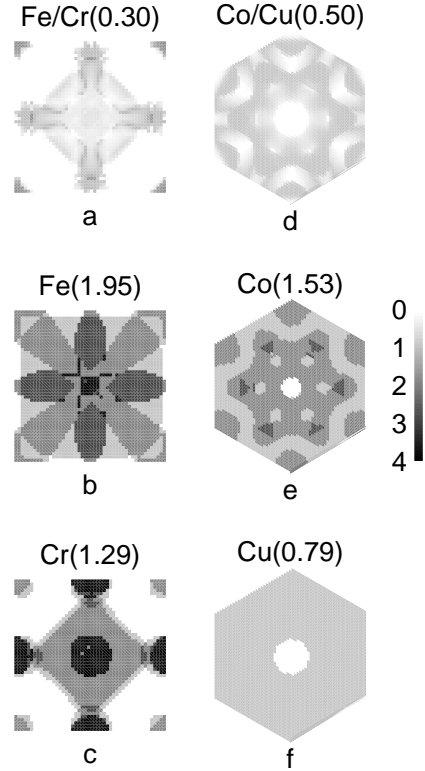}
\caption{ 
Number of propagating channels in first Brillouin Zone.
(a), (b) and (c) are for the majority spin of a clean Fe/Cr
interface, bulk Fe, and bulk Cr,(d), (e) and (f) are for minority
spin electrons of a clean Co/Cu interface, bulk Co, and bulk Cu,
respectively. The total number of propagating channels per unit
cell is given in brackets on top of the figures. The grayscale
interpolates the number of propagating channels per $k_{\|}$-point
between zero (white) and four (black). }
\label{Fig2}
\end{figure}

The Fe/Cr (non-magnetic) interface calculation is performed for the bcc(100)
orientation, which is the low index orientation with the largest
spin-asymmetry\cite{Stiles00}. We used a lattice constant of 2.87
\AA . To model the interface roughness a $6 \times 6$ lateral supercell was
used. The uncertainty from configuration averaging is less than 10\%.

Whereas for Co/Cu the majority-spin band structures were well matched,
for Fe/Cr the situation is reversed and it is the minority-spin
electronic structures which match well.
Using Eq.~(\ref{Rint}), the interface resistance is
${\rm 2.82 ~f\Omega m^{2} }$ and
${\rm 0.50 ~f\Omega m^{2} }$ for majority and minority spin,
respectively, so that the Fe/Cr interface has a negative
spin-asymmetry, opposite to that of bulk Fe.
The effect of disorder is to suppress the interface asymmetry rather
than enhance it. 
As was the case for the Co/Cu majority spins, interface disorder has
only a small effect on the well-matched minority spin channel.
For the majority spin channel however, the transmission probability
for a clean interface is very low due to a large band mismatch.
For a disordered interface, the ballistic contribution to the
conductance can only decrease by a small amount but the diffuse
component increases enormously leading to a large net increase in the
transmission.
3\% Fe impurities in the first Cr layer (or 3\% Cr in the first Fe
layer) increase the transmission probability by more than 10\%.
Two interdiffused atom layers suppress the spin asymmetry and the MR
efficiently - the
interface resistances resulting from two 50-50 interface alloy layers are
${\rm 0.99 ~f\Omega m^{2} }$ and
${\rm 0.50 ~f\Omega m^{2} }$ for majority and minority spin
respectively. {\em Vice versa}, the interface quality is much
more critical for a large CPP-MR in the Fe/Cr than in the Co/Cu system.


The qualitative difference between Fe/Cr and Co/Cu can be understood using
Figs.~\ref{Fig2}(a-c) and Figs.~\ref{Fig2}(d-f) where we show as a
function of ${\bf k_{\|}}$ the number of majority-spin propagating
channels for the Fe/Cr interface, bulk Fe, and bulk Cr,
respectively, and the number of minority-spin propagating channels
for the Co/Cu interface, bulk Co, and bulk Cu, all in the first BZ.
For the Co/Cu (111) interface minority-spin states, the Fermi
surfaces of both Co and Cu occupy a large part of the 2D BZ so that there
are a large number of states with the same ${\bf k_{\|}}$
in both materials which can, in
principle, propagate in the absence of disorder.
Summing over all ${\bf k_{\|}}$, however, the transmission
probability of states (coming from Cu) is only about 60\%; the
character of the bulk states on either side is such that they match
poorly. Defect scattering tends to reduce the transmission probability
and thus {\em increases} the interface resistance of the Co/Cu minority spin
channel. 
On the other hand, we can identify two mechanisms by which
interface disorder {\em decreases} the interface resistance in the
Fe/Cr majority-spin channel by a factor 3.
Majority spin electrons with small ${\bf k_{\|}}$ are almost
completely reflected at the Fe/Cr specular
interface because the electronic states on both sides of the interface do
not match well. 
Defect scattering is found to increase the transmission of these electrons
strongly.
Furthermore, for ${\bf k_{\|}}$ outside of this central area, there are no
propagating
states on the Cr side. Propagating modes in Fe with larger ${\bf k_{\|}}$,
which are totally
reflected at the specular interface, can be scattered diffusely into the
center of the BZ
where many states are available in Cr.

In summary, we have studied the interface resistance of Co/Cu, Fe/Cr
and Au/Ag interfaces. Depending on the system, interface disorder
can increase or decrease the interface resistance.
For some interfaces such as Au/Ag and Co/Cu, the band mismatch at an
interface is responsible for most of the interface resistance.
For other systems such as Fe/Cr, the interface resistance
can dramatically depend on the interface perfection.
For Fe/Cr interface, the majority-spin interface resistance is
reduced as much as 70\% by interface disorder.
Interface disorder enhances the spin asymmetry in the Co/Cu system
but decreases it for Fe/Cr.
In the systems considered, the diffuse scattering arising from interface
disorder breaks the
phase coherence in high resistance spin channels, but not necessarily for
the low
resistance spin channels.

This work is part of the research program for the ``Stichting voor
Fundamenteel Onderzoek der Materie'' (FOM), which is financially
supported by the ''Nederlandse Organisatie voor Wetenschappelijk
Onderzoek'' (NWO).  This study was supported by the NEDO joint
research program (NTDP-98), the Grant Agency of the Czech Republic
(202/00/0122), and the Grant Agency of the Academy of Sciences of
the Czech Republic (A1010829). We acknowledge benefits from the TMR
Research Network on ``Interface Magnetism'' under contract No.
FMRX-CT96-0089 (DG12-MIHT).

\end{document}